# Electric field control of deterministic current-induced magnetization switching in a hybrid ferromagnetic/ferroelectric structure

Kaiming Cai[1†], Meiyin Yang[1†], Hailang Ju[2], Kevin William Edmonds[3], Baohe Li[2], Yu Sheng[1], Bao Zhang[1], Nan Zhang[1], Shuai Liu[2], Yang Ji[1], Houzhi Zheng[1] and Kaiyou Wang[1*]

1. SKLSM, Institute of Semiconductors, CAS, P. O. Box 912, Beijing 100083, China
2. Department of Physics, School of Sciences, Beijing Technology and Business University, Beijing 100048, China
3. School of Physics and Astronomy, University of Nottingham, Nottingham NG7 2RD, United Kingdom
† These authors contributed equally to this work.
* Correspondence and requests for materials should be addressed to K. W. (e-mail: kywang@semi.ac.cn).

**All-electrical and programmable manipulations of ferromagnetic bits are highly pursued for the aim of high integration and low energy consumption in modern information technology[1-3]. Methods based on the spin-orbit torque switching[4-6] in heavy metal/ferromagnet structures have been proposed with magnetic field[7-15], and recently are heading toward deterministic switching without external magnetic field[16,17]. Here we demonstrate that an in-plane effective magnetic field can be induced by an electric field without breaking the symmetry of the structure of the thin film, and realize the deterministic magnetization switching in a hybrid ferromagnetic/ferroelectric structure with Pt/Co/Ni/Co/Pt layers on PMN-PT substrate. The effective magnetic field can be reversed by changing the direction of the applied electric field on the PMN-PT substrate, which fully replaces the controllability function of the external magnetic field. The electric field is found to generate an additional spin-orbit torque on the CoNiCo magnets, which is confirmed by macrospin calculations.**

In heavy metal (HM)/ferromagnet (FM) structures with large perpendicular inversion asymmetry, an applied electric current can generate a spin current which provides an effective way to control the magnetization[1,2,20,21]. An advantage of these switching mechanisms is the simplicity of the film structure, compared to conventional magnetic random access memory (MRAM) elements in which a magnetic polarization layer is generally required to generate the spin-polarized electrons. Magnetization with perpendicular anisotropy could be switched to a fixed direction under an in-plane magnetic field[5,6,10,14], or otherwise switched to a random up or down direction without a magnetic field[22]. More recently, a lateral wedge oxide[16,19,23] or anti-ferromagnetic layer[24] has been used to induce the deterministic switching without external magnetic field. Deterministic switching was also realized in structures with a tilted magnetic easy axis[17,18]. However the deterministic switching due to structural asymmetry is not controllable once deposited. Thus controllable and reliable means to manipulate the magnetization all-electrically by spin-orbit torque (SOT) are highly demanded.

Recent experimental and theoretical work suggest that the deterministic switching of magnets under an in-plane external magnetic field ($H_x$) applied along the current direction is due to the field-like torque, which creates an unequal energy barrier for switching the magnetization between up and down. Deterministic switching in the absence of external magnetic field becomes possible owing to the spin-orbit coupling of the motion of electrons under electric field, producing an effective magnetic (spin-orbit) field acting on the spins[20]. As a result, the spin-orbit field also exerts a torque on the local magnetization[15,26,27]. Here, we utilize the electric field produced in PMN-PT substrates to control the SOT in the ferromagnetic layer in the hybrid ferromagnetic/ferroelectric structure, and realize the magnetization switching without an external magnetic field. Moreover, the deterministic switching direction can be controlled by reversing the electric field, thereby enabling a programmable function. The discovery and design of the electric field controlled spin-orbit torque deterministic switching holds great potential application for spintronic memory and logic devices.

Experiments were carried out on sputter-deposited Pt(4 nm)/CoNiCo(0.4/0.2/0.4 nm)/Pt(2 nm) multilayer films on PMN-PT substrates as the hybrid ferromagnetic/ferroelectric structure (inset of Fig. 1a). Details of sample preparation are presented in the Methods. A symmetric structure with top and bottom Pt layers was chosen, which was intended to minimize the Rashba effect[8]. The bottom Pt layer was patterned and etched into a cross-shaped structure, and the CoNiCo magnetic layer film was etched into a pillar array with 3 μm in diameter at the center of the Hall cross (as shown in Fig. 1a). With applied out-of-plane magnetic field, the anomalous Hall resistance ($R_H$) follows a square-shaped magnetic hysteresis loop as shown in Fig. 1b, indicating that the magnetization easy axis is along the $z$ direction. No obvious change of magnetic hysteresis loops were observed with different voltages (0, ±100, ±200, ±500 V) applied on the substrate (Supplementary Fig. S3), suggesting that the magneto-elastic coupling between the substrate and the CoNiCo ferromagnet is very weak and can be ignored.

Current-induced switching under in-plane magnetic field $H_x$= ±3 Oe is shown in Fig. 1c. The opposite $H_x$ results in the current-induced switching through opposite path, which is consistent with previous reports[5,7,10]. To our surprise, in the absence of external magnetic field, we observed that the magnetization was deterministically switched to −$M_z$ (+$M_z$ for "up" and −$M_z$ for "down") for both forward and backward current sweeping directions (Fig. 1d) with the critical current ~ 1.5 mA. The unidirectional switching behaviors in our experiments are well reproducible. However, the magnetization of a similar structure on Si substrate did not switch deterministically without magnetic field[10,22]. Thus, the unidirectional switching is attributed to the PMN-PT substrate.

To investigate the influence of PMN-PT substrate on the spin-orbit torque switching, we applied a voltage of ±500 V across the substrate in a direction parallel to the current direction of the Hall bar (Fig. 2a). The voltage was removed during the current-induced switching measurements. A clockwise loop of magnetization switching was observed after applying +500 V on the substrate, as shown in Fig. 2b, where the current sweeping from positive to negative favors the down magnetization and that from negative to positive favors the up direction. The loop is reversed to anticlockwise under −500 V in Fig. 2c. Thus, the controllable deterministic switching of the CoNiCo magnet was achieved by electric fields applied on PMN-PT substrates. Using the controllable switching phenomenon, we could switch the magnetization of the CoNiCo magnet between up and down using current pulses after applying 500 V voltage across the substrate, as shown in Fig. 2d. This demonstrates a series of writing and reading of information by injecting $\pm I$ pulses, which could be used as a MRAM.

Further investigations were focused on the origin of the voltage-controlled deterministic switching. Figure 3a shows the spin distribution due to the spin Hall effect of Pt with current injected into the Pt leads. If only spin Hall effect exists, the spin density can switch the magnetization up or down randomly with equal probability. However, there is an electric potential difference across the PMN-PT substrate under the device when injecting current to the Hall bar, resulting in an external electric field $E_{1x}$ which could align the electric dipole of the PMN-PT along $x$-direction. The $E_{1x}$ was estimated to be ~ 1.6 kV/cm with the voltage of 4 V around the magnetic dot area, which is sufficiently large enough to polarize the electric dipole (Supplementary S2). The polarization of PMN-PT induces an opposing electric field $E_{2x}$. The internal electric field of PMN-PT was largely reduced due to the polarization with $E_{int}=E_{1x}-E_{2x}$, so the potential slope in PMN-PT is smaller than that of Pt as shown in Figs 3b and d. The differences of electrical potential between the Pt and PMN-PT interface results in a gradient perpendicular electric field $E_p$, which decreases from left to right under positive voltage. With negative voltage applied, the $E_p$ reverses its direction as shown in Fig. 3d. The penetration length of $E_p$ into the Pt layer was estimated to be ~1 nm by Debye-Huckel approximation (Supplementary S4), which is approximately a quarter of the total thickness of Pt. The conduction electrons near the Pt/PMN-PT interface experience the spin-orbit coupling field due to the perpendicular electric field $E_p$, which results in non-equal spin-up and spin-down densities as shown in Fig. 3c. For $-E_p$, the spin-down electrons are more favorable at the interfaces between the PMN-PT/Pt as is shown in Fig. 3c. The total spin density $J_S$ at the Pt/CoNiCo interface is the superposition of the two effects--the spin Hall effect and the spin-orbit coupling due to $E_p$. Figures 3e and f present the spin distribution under $+E_p$ and $-E_p$, respectively, with the positive current injecting along Pt, which produce opposite spin density gradients between the CoNiCo and lower Pt interfaces. There are four conditions for spin density distribution as illustrated in Fig. 4a.

The gradient of spin densities or magnetization can induce an extra torque[28] during the current-induced magnetic switching. Based on the analysis of the gradient spin current density in our experiments, we derived a formula (Supplementary S5) which shows that the gradient spin density exerts an additional torque of $\tau_\mathbf{n} \approx -c\mathbf{M} \times (\partial \mathbf{J}_s/\partial x)$, with a constant $c$ and the spin current density $\mathbf{J}_s$. The $\tau_\mathbf{n}$ either is parallel or anti-parallel to the spin transfer torque depending on the sign of the spin density gradient as was illustrated in Figs 3e and f. Thus, the LLG equation for current induced switching under external electric field can be rewritten as:

$$\frac{\partial \mathbf{M}}{\partial t} = -\gamma \mathbf{M} \times \mathbf{H}_{eff} + \frac{\alpha}{M_s}\mathbf{M} \times \frac{\partial \mathbf{M}}{\partial t} - c\mathbf{M} \times \frac{\partial \mathbf{J}_s}{\partial x} \qquad (1)$$

The damping torque and the spin transfer torque are combined together in the second term of the equation as they have the same form[29]. To explain how the torque $\tau_n$ rotates and switches the magnetic orientation of a perpendicular magnet, the new equilibrium orientations of magnetization were analyzed with a macrospin model[4] by solving the total torque and balance equation with $\vec{\tau}_{tot} = \vec{\tau}_{ST} + \vec{\tau}_{an} + \vec{\tau}_n = 0$ (details in Supplementary S5). The calculation used a constant $\partial \mathbf{J}_s/\partial x$ for the spin current density which increased approximately linearly or decreased linearly due to the constant gradient $E_p$. Figure 5a shows that the positive voltage ($b=(\partial J_s/\partial x)/B_{an}^0 > 0$) induces a clockwise equilibrium $M_z$ loop and negative voltage ($b=(\partial J_s/\partial x)/B_{an}^0 < 0$) leads to the anticlockwise loop, under the spin transfer torque $a=\tau_{ST}^0/B_{an}^0$. The term b > 0 occurs when the perpendicular field $E_p < 0$ due to the applied positive voltage. The calculation result agrees with the experiment for the loops under external voltages in Fig. 2a.

As expected, the unidirectional magnetization switching behavior with positive and negative directions of current in Fig. 1d can be also explained by the model. The current applied across the Pt results in a voltage applied on the PMN-PT substrate, producing a sign change of the voltage while sweeping the current direction. Thus, considering the electric field induced by the applied current, the switching behavior follows the switching loop in Fig. 4c at positive current, but changes to follow the loop in Fig. 4b when reversing the current to negative. Therefore, unidirectional switching of $-M_z$ was observed.

The voltage controlled deterministic switching of perpendicular magnetic bits in this study can be used as a non-volatile magnetic memory with programmable functionality, such as encrypted memory. Multiple functional logic with only one or two magnetic bits can be realized using the same device because of the programmable magnetization switching by the voltage, which will greatly advance the information technology with simple design.

**Methods**

**Device fabrication.** The stack structure of Pt/CoNiCo/Pt was deposited on a single crystal (001) PMN-PT substrate by d.c. magnetron sputtering at room temperature. The deposition rates were 0.075 nm s$^{-1}$ for Pt, 0.047 nm s$^{-1}$ for Co and 0.042 nm s$^{-1}$ for Ni at pressure of 0.5 Pa, respectively. The samples were subsequently patterned into an array of Hall bar devices by standard EBL and etching techniques. The size of the Hall bars was fixed at 8 μm×25 μm and a nano-pillar with 3 μm in diameter in the center of the Hall bar.

**Hall measurements.** For transport studies, Keithley meters and lock-in amplifiers were used in our measurements. A Keithley 2602 current source and a Keithley 2182A nanovoltmeter were used in the d.c. Hall voltage measurements. All measurements were carried out at room temperature. More than 20 devices with two different thickness of bottom Pt layers were tested in the same batch and similar behaviors were also observed in other devices.

**Numerical simulations.** A single CoNiCo macrospin represented the uniform magnetization *M* of our switching feature, and we used a fourth-order Runge-Kutta integration of the Landau-Lifshitz-Gilbert (LLG) equation. The calculations were carried out assuming finite temperature (300 K) and sweeping either *I* or *H*.

**Acknowledgements**

This work was supported by "973 Program" No. 2014CB643903, NSFC Grant No. 61225021, 11474272 and 11174272.

**Author contributions**

K. W. and M. Y. conceived and designed the experiments. H. J., S. L. and B. L. provided the thin magnetic films. K. C. and M. Y. fabricated and measured the devices. K. W., M. Y. and K. C. performed the theoretical analysis and modeling. B. Z., Y. S., and N. Z. assisted in measurement and processed the data. K. W. E and Y. J. and H. Z. discussed the results and commented on the manuscript. K. C., M. Y. and K. W. wrote the manuscript. All authors revised the manuscript.

**Additional Information**

The authors declare no competing financial interests. The details of the characterizations and calculations were shown in the supplementary information. Correspondence should be addressed to K. W. (kywang@semi.ac.cn).

**Competing financial interests**

The authors declare no competing financial interests.

**Figures and figure captions**

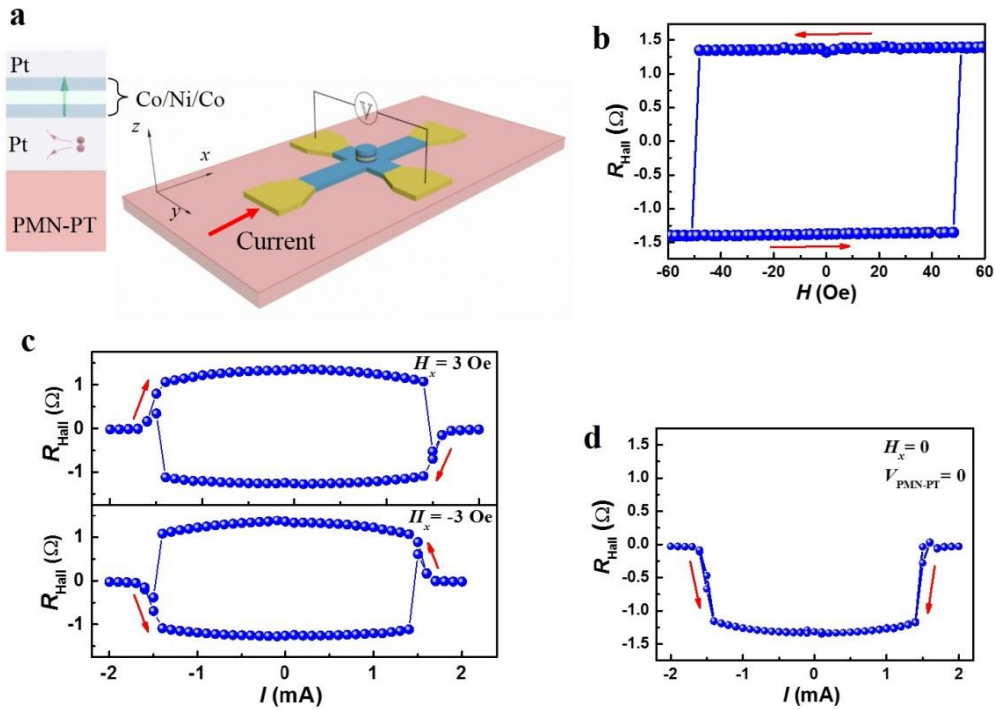

**Figure 1 | Schematic of device structures and current switching measurements of the devices. a**, The structure of the device. The Hall cross in lateral direction is 8 $\mu$m wide. The transverse direction is 2 μm wide for the Hall voltage detection. The CoNiCo/Pt dot is a circle with 3 μm in diameter. Inset is the typical cross-section of the deposited stack, with the green arrow showing the magnetic easy axis of the Co/Ni/Co. **b**, The anomalous Hall resistance loop of the device under a perpendicular magnetic field. **c**, The current induced magnetization switching probed by anomalous Hall effect under in-plane magnetic field with $H_x$= + 3 and − 3 Oe. **d**, The current induced deterministic switching without in-plane magnetic field.

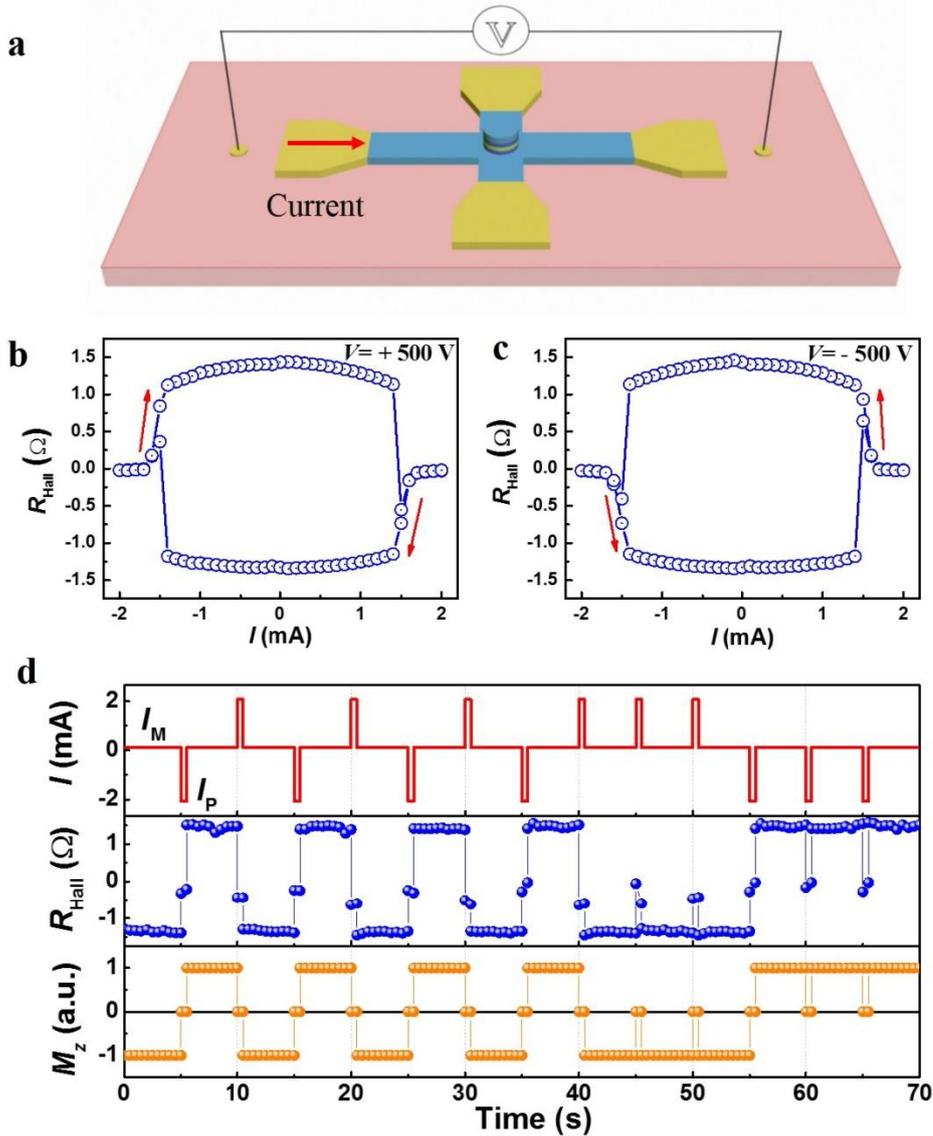

**Figure 2 | Electrical controllable deterministic magnetization switching by current pulses without magnetic field. a**, The schematic of the measurement set-up. The voltage is applied on the PMN-PT substrates with two terminals nearby the two electrodes which were used to inject current along *x*-direction. The voltage was removed during the measurements. **b & c**. The current-induced magnetization switching after the polarization with + 500 V and – 500 V on PMN-PT substrate. **d**, The deterministic magnetization switching by a series of current pulses. A small current $I_M$ (~ 0.1 mA) was applied to measure the Hall resistance to distinguish the magnetization state.

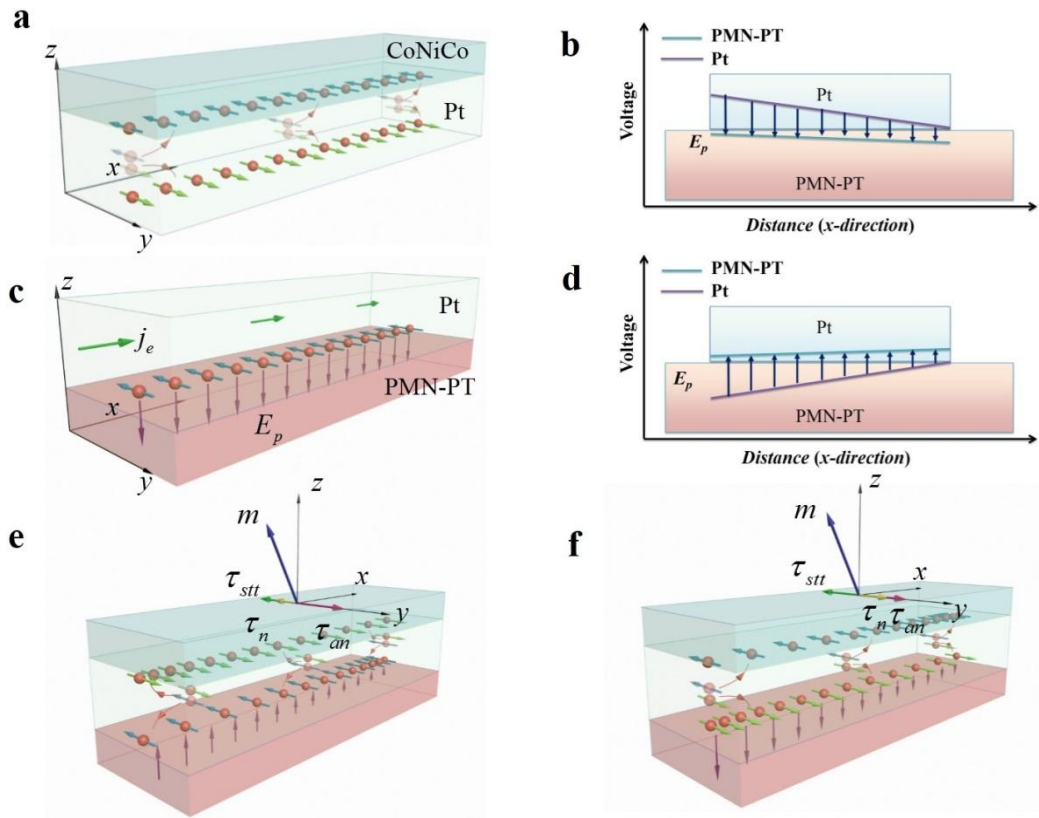

**Figure 3 | The origin of the deterministic switching. a,** The spin distribution in Pt when injecting a current due to the spin Hall effect. **b & d,** The electric field at the interface between the PMN-PT substrate and the bottom Pt layer during the positive current $+I$ and negative current $-I$ current injection. A potential difference is formed in z-direction due to the polarization of substrate. **c,** the spin distribution at the Pt/PMN-PT interface due to the spin-orbit coupling by the perpendicular electric field. **e & f,** With $E_p > 0$ and $E_p < 0$, the spin-orbit torque of the illustration of CoNiCo magnets due to the superposition of the spin Hall effect and the spin-orbit coupling.

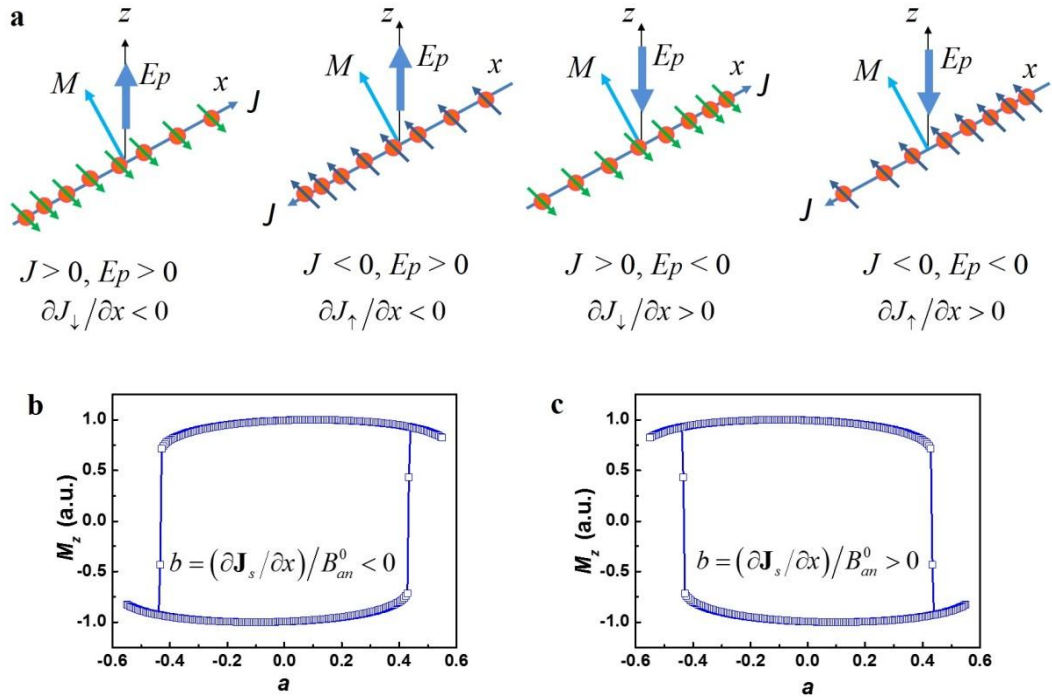

**Figure 4 | Spin current density distribution and the magnetization switching. a,** Four conditions of spin current density distribution at the Pt/CoNiCo interface depending on current direction $J$ and perpendicular electric field $E_p$. **b,** The switching loop calculated with $\partial \mathbf{J}_s/\partial x < 0$, which is associated with the negative voltage applied on the PMN-PT substrate. **c,** The switching loop calculated with $\partial \mathbf{J}_s/\partial x > 0$, which is associated with the positive voltage applied on the substrate.